# Effect of oxygen off-stoichiometry on coupled structural and magnetic phase-transitions in La$_{0.15}$Sr$_{0.85}$MnO$_{3-\delta}$ ($\delta$ = 0.02, 0.14)

Aga Shahee, R.J.Chaudhari, R.Rawat, A.M.Awasthi and N. P. Lalla

UGC-DAE Consortium for Scientific Research, University campus, Khandwa road Indore, India- 452017

## Abstract

Effect of oxygen off-stoichiometry on the structural and magnetic properties of La$_{0.15}$Sr$_{0.85}$MnO$_{3-\delta}$ ($\delta$=0.02, 0.14) has been studied employing low-temperature x-ray diffraction, calorimetry, electrical-transport and magnetization. The nearly stoichiometric composition under goes first order phase-transition from cubic to tetragonal P4/mmm at 238K accompanied by magnetic and electrical phase-transitions. On the contrary the off-stoichiometric composition does not show any transition and remain cubic Pm-3m down to 80K. Off-stoichiometry induced oxygen-vacancy seems to suppress the cooperative JT distortion and possibly stabilize a G-type antiferromagnetic phase with localized canted spins. Thus we demonstrate that controlling oxygen stoichiometry is a way of structure/property engineering in electron doped like manganites.

**Keywords:**   **A.** Manganite, **D.** Phase-transition, **D.** Cooperative JT-Distortion, **E.** Low Temperature x-ray Diffraction

**Corresponding Author:**  Niranjan Prasad Lalla

**E-mail ID:** nplallaiuc82@gmail.com

**Tel./Fax Number** : +91-0731-2463913   /   +91-0731-2462294

**Introduction:**

Extensive studies probing the properties of mixed valent manganites with the perovskite structure have been carried out for almost 60 years, which have revealed exciting physical and structural properties. These properties are found very sensitive to the type and strength of lattice distortions[1,2], the filling of one-electron band and its width[3], internal strain[4,5], A-site disorder[6,7], and site vacancy as well as external perturbations, e.g., magnetic field and pressure[8]. The competition between these parameters gives rise to the bicritical or multicritical states with respect to one or more control parameter(s) [9, 10]. Thus manganites offer a degree of chemical flexibility which permits us to control and manipulate its structural, electronic and magnetic states. The competition of ferromagnetic metallic state with other orbital-charge-spin ordered states; typically the CE-type antiferromagnetic (AFM) charge-ordered and orbital-ordered insulating state has led to technologically important physical phenomena like colossal magnetoresistance [11,12] and dense granular magnetoresistance [13] and also promises manganite to be a candidate for spintronics applications. In this context, another important control parameter appears to be oxygen stoichiometry. Oxygen vacancies in perovskite oxides are fundamentally interesting and technologically important [14]. Oxygen vacancy defects are frequently found in $ABO_3$ and significantly alter their properties. For example, fatigue in the switching of ferroelectric non-volatile RAM devices[15], degradation of the dielectric and piezoelectric properties[16], induction of ferromagnetism and metal-insulator transition[17], affecting the spin and charge ordered states[18] and the memory effect[19]. Such effects of oxygen off-stoichiometry on the structural, magnetic and electrical properties of $LaMnO_{3-\delta}$[20-22], A-site deficient $La_{1-x}MnO_{3\pm\delta}$[22-24] and hole doped $La_{1-x}Sr_xMnO_{3\pm\delta}$ (x<0.5)[25] and $La_{1-x}Ca_xMnO_{3\pm\delta}$ (x=0.1& 0.3) [26] samples have been extensively studied.

However, for electron doped like $La_{1-x}Sr_xMnO_3$ (x > 0.5) where A, C & G-type AFM insulating states get stabilize and compete with each other, only few experimental reports on the effect of oxidation state of Mn [27,28] and oxygen non-stoichiometry are available[29-31]. In this article we report the effect of oxygen off-stoichiometry on the behavior of coupled structural and magnetic phase-transitions in $La_{0.15}Sr_{0.85}MnO_{3-\delta}$ perovskite and present a completely different way to manipulate and control the structural, electronic and magnetic states of an electron doped like manganite $La_{0.15}Sr_{0.85}MnO_{3-\delta}$.

**Experimental:**

$La_{0.15}Sr_{0.85}MnO_{3-\delta}$ ceramics were prepared following the conventional solid state route. Well mixed stoichiometric mixture of high purity (99.99%) ingredients $La_2O_3$, $SrCO_3$, and $MnO_2$ was calcined at $1000^oC$ for 24 hrs in alumina crucible. The calcined powder was reground for ~2 hours and then pelletized and sintered at 1400°C for 60 hours. After completion of the sintering process the furnace was slowly cooled at the rate of 2°/min. The sintered pellets where divided into two batches, one batch was left as such and here after will be termed as air-sintered LSMO and the second batch was annealed under argon at $1000^oC$ for 6hrs and then slow cooled at the rate of 2°/min. The crushed powder of both LSMO samples was subjected to XRD followed by Rietveld refinement [32] using Pm-3m space-group. Absence of any unaccounted peak revealed that the samples are single cubic phase. EDAX analysis revealed that atomic percent ratio of the elements (La, Sr and Mn) is within ±1% of the atomic percent of the synthesized composition. Oxygen stoichiometry was determined using idiometric titration with error estimated around ±0.01. It showed that the as-sintered $La_{0.15}Sr_{0.85}MnO_{3-\delta}$ was nearly stoichiometric with δ=0.02±0.01 and the argon-annealed sample was off-stoichiometric with δ=0.14±0.01. The well characterized LSMO samples were then subjected to low-temperature structural studies from 320K to 80K, employing a XRD setup (Rigaku D-max) equipped with a $LN_2$ based cryostat and mounted on a x-ray generator operated at 50 KV and 200 mA. The LT-XRD data was Rietveld refined using the FullProf [32]. To corroborate the structural phase-transition studies with the occurrence of latent-heat and change in heat-capacity ($C_p$), modulated differential scanning calorimetry (MDSC) was carried out using TA-Instruments (Model 2910) with cooling/heating rates of $2^oK/min$ and modulation parameters of 1K:1Hz. Electrical-transport measurements down to ~50K were done using 4-probe resistance-vs-temperature (R-T) measurements setup (Keithley nano-voltmeter model-182 and constant-current source 2400 and Lakeshore model DRC-93CA). Zero-field cool (ZFC) and field cool (FC) magnetization (M-T) measurements were done at 100Oe and 7T fields from 470K down to 50K using SQUID-VSM setup (Quantum Design).

**Results and Discussion:**

Figs.1 (a,c) and (b,d) show the Rietveld refined XRD patterns of the air-sintered $La_{0.15}Sr_{0.85}MnO_{2.98}$ ($\delta=0.02$) and argon-annealed $La_{0.15}Sr_{0.85}MnO_{2.86}$ ($\delta=0.14$) samples taken at 320K and 80K respectively. The room temperature (RT) XRD data (a,b) were best fitted with cubic space-group Pm-3m. From Figs.(b,d) it can be seen that $La_{0.15}Sr_{0.85}MnO_{2.86}$ ($\delta=0.14$) remains cubic down to 80K whereas the stoichiometric $La_{0.15}Sr_{0.85}MnO_{2.98}$ ($\delta=0.02$) under goes a cubic to tetragonal phase transition. The tetragonal phase was found to best fit with P4/mmm [33] with La/Sr at (1a), O at (1b & 1c) and Mn at (1d) Wyckoff positions. The results of Rietveld refinement of the XRD data taken at various temperatures between 320K and 80K is summarized in Figs.2 (a,b). It is clear from Fig.2a that the RT lattice parameter of the cubic phase of off-stoichiometric $La_{0.15}Sr_{0.85}MnO_{2.86}$ is greater than that of the stoichiometric $La_{0.15}Sr_{0.85}MnO_{2.98}$. From Fig.2a it is clear that the oxygen deficient $La_{0.15}Sr_{0.85}MnO_{2.86}$ sample does not show any structural phase-transition and remain cubic down to 80K. It simply shows a monotonous decrease of the cubic lattice parameter due to thermal contraction during cooling. The stoichiometric $La_{0.15}Sr_{0.85}MnO_{2.98}$ sample shows a cubic (Pm-3m) to tetragonal (P4/mmm) phase-transition at ~238K with phase coexistence region between 240K-220K, see Fig.2a. Fig.2b shows the occurrence of discontinuous change in the unit-cell volume during cubic-tetragonal phase-transition. Occurrence of phase-coexistence and discontinuous change in the unit cell volume are the evidence of the first-order phase-transition (FOPT). To vividly show the absence of cubic-tetragonal phase-transition for the oxygen deficient $La_{0.15}Sr_{0.85}MnO_{2.86}$ cubic phase the splitting behaviour of a reasonably high order degenerate reflection (330)/(411) of both the phases was monitored. Corresponding to tetragonal transformation of $La_{0.15}Sr_{0.85}MnO_{2.98}$ cubic phase across $T_s$= 238K, the occurrence of splitting of the degenerate cubic reflection (330)/(411) into (303),(114),(330) & (411) can be clearly seen in Fig.2c. On the contrary the degenerate reflection (330)/(411) of $La_{0.15}Sr_{0.85}MnO_{2.86}$ cubic phase does not show any such splitting down to 80K, see Fig.2d. The c/a ratio of 1.017 of the pseudo perovskite cell of the tetragonal phase and the Mn-O-Mn bond length along c-axis were found to expand. This was analyzed to be due to cooperative JT-distortion of the $MnO_6$ octahedra resulting in ferro-orbital ordering of $3d_{3z^2-r^2}$[1]. This will result in a coupled C-type anti-ferromagnetic ordering across the cubic-tetragonal phase-transition [27, 28].

Fig.3 shows the variation of total heat-flow (latent-heat) and specific heat-capacity ($C_p$) as a function of cooling and heating for (a) $La_{0.15}Sr_{0.85}MnO_{2.98}$ and (b) $La_{0.15}Sr_{0.85}MnO_{2.86}$. It can be seen that the stoichiometric sample $La_{0.15}Sr_{0.85}MnO_{2.98}$ clearly shows exothermic (cooling) and endothermic (heating) peaks in the total heat-flow and $\Delta$-transition peaks in the $C_p$-T variation. The exothermic and endothermic peaks correspond to the latent-heat associated with the cubic-tetragonal and coupled C-type AFM phase-transitions. The presence of latent-heat shows the first-order nature of this phase-transition. From Fig.3b it is clear that off-stoichiometric sample neither shows the exothermic/endothermic peaks in the total heat-flow nor the $\Delta$-transition like peaks in $C_p$-T. This confirms the absence of phase-transition in off-stoichiometric sample. Thus the MDSC results fully support the LT-XRD result, which shows complete suppression of cubic-tetragonal phase-transition in off-stoichiometric $La_{0.15}Sr_{0.85}MnO_{2.86}$.

The oxygen off-stoichiometry is also expected to affect physical properties. The stoichiometric $La_{0.15}Sr_{0.85}MnO_{2.98}$ shows a step-like feature in the $(R_T/R_{300})$-T at temperature $(T_s/T_N)$ of 238K, see Fig.4. The step feature in R-T corresponds to cubic-tetragonal structural phase-transition coupled to C-type AFM phase-transitions [28]. On the other hand off-stoichiometric sample $La_{0.15}Sr_{0.85}MnO_{2.86}$ does not show any step-like feature but a monotonous increase in the R-T. This further confirms the suppression of phase-transition in $La_{0.15}Sr_{0.85}MnO_{2.86}$. The $(R_T/R_{300})$-T data of off-stoichiometric LSMO shows that the relative rate of increase of resistance with decreasing temperature is higher than that of the stoichiometric LSMO. This clearly exhibits the emergence of characteristically different electrical properties for the off-stoichiometric $La_{0.15}Sr_{0.85}MnO_{3-\delta}$.

Fig.5 shows 100Oe and 7T ZFC and FC M-T data. The 100Oe M-T shows paramagnetic to weak-ferromagnetic (WF) like transition around 355K and also bifurcation of ZFC and FC variation. Using Curie-Weiss plot and M-H behavior, data not shown, the WF behavior is analyzed to be due to appearance of some minor regions with ferromagnetic interaction arising from Griffith-phase like singularity [34-36]. Occurrence of Griffith phase in antiferromagnetic manganites is well reported [37, 38]. Keeping in view that the observed WF behavior is equally present for both types of samples the commonly existing A-site substitutional disorder of La and Sr appears to be its origin. The off-stoichiometry induced change in the major magnetic phase is seen as the suppression of the sharp C-type AFM phase-transition at 238K but it is over whelmed by the WF feature in the low-field M-T measurement. According to Goodenough [1] and early experimental reports [2] across a JT distortion induced cubic-tetragonal phase-transition a C-type AFM phase gets stabilize as seen currently for $La_{0.15}Sr_{0.85}MnO_{2.98}$. The drastic effect of off-stoichiometry on the major magnetic phase realized in the 7T M-T data, where WF effect is totally suppressed. It can now be seen that oxygen off-stoichiometry has nearly suppressed the sharp C-type AFM phase-transition but leaving behind a broad kink like feature whose possible origin is discussed below.

Low temperature XRD, MDSC, R-T and M-T measurements very clearly approve that the oxygen stoichiometric $La_{0.15}Sr_{0.85}MnO_{2.98}$ sample under goes a first-order cubic to tetragonal phase transition, with c/a =1.017 accompanied with a C-type AFM ordering[27,28] while the oxygen off-stoichiometric $La_{0.15}Sr_{0.85}MnO_{2.86}$ remains Pm-3m cubic. From Fig.2b it is also evident that the unit-cell volume of $La_{0.15}Sr_{0.85}MnO_{2.86}$ is greater than the unit cell volume of $La_{0.15}Sr_{0.85}MnO_{2.98}$ at all temperatures. This is because oxygen deficiency increases the $Mn^{3+}$ concentration whose ionic radius is larger than $Mn^{4+}$ and therefore in off-stoichiometric sample effective Mn size will increase. The observed facts appears to be against the known phase diagram for manganites, where structural stability has been realized for higher $Mn^{3+}$ ion concentration and according to which the oxygen deficient $La_{0.15}Sr_{0.85}MnO_{2.86}$ sample (with 43% $Mn^{3+}$) should stabilize in A-type AFM phase at low temperatures. This discrepancy can be under stood as in the following.

Cooperative JT-distortion (CJTD) in $ABO_3$ perovskite appears due to cooperativity between the local octahedral units $BO_6$ known as "vibrons" [39], where B is a JT active-ion. It has been shown experimentally [40, 41] that for a CJTD to occur a threshold concentration of JT active units and connectivity between them is necessary. For example in $(Fe_{1-x}Mn_x)_3O_4$ the threshold value of x for tetragonal distortion is 0.4[40]. In both of our cases minimum concentration condition is fulfilled, so it appears that off-stoichiometry induced oxygen-vacancy (OV) is the origin of disconnectivity between JT active-ion and suppression of cooperative JT-distortion. It appears that off-stoichiometry induced oxygen-vacancy (OV) is the origin of the effective decrease of the threshold value of JT-active octahedral units in the present case. Recently through density function calculation on SrTiO$_3$ Lin et. al [42] have shown that lowering of the local symmetry from cubic (O$_h$) to tetragonal (C$_{4v}$) around a OV in transition metal (TM) perovskite oxides induces direct on-site coupling between the 3d$_{3z2-r2}$ and 4s,4p orbitals of the TM atom adjacent to the OV. Due to OV induced imbalance of direct coulomb repulsion between e$_g$-charge and O$^{2-}$ ions the charge density lobs of 3d$_{3z2-r2}$ orbital are localized about the OV along the O$^{2-}$-TM-OV-TM-O$^{2-}$ axis. Such a situation at one hand introduces a direct hopping between the two TM atoms lying just across the OV but on the other hand will suppress local JT-distortion around OV-site and also reduces indirect hopping between the next nearest TM atoms. Thus the e$_g$-charge associated with 3d$_{3z2-r2}$ orbital of TM-OV-TM (Mn$^{3+}$-OV- Mn$^{3+}$) unit will be highly localized and will reduce local symmetry from cubic to C$_4^V$ instead of cubic to JT-distorted tetragonal [42]. Thus will act as a blocking point for the e$_g$-charge hopping along the O$^{2-}$-TM-O$^{2-}$-TM-O$^{2-}$ chain and cooperative JT-distorted tetragonal. Since pairing of TM atoms through the formation of (TM-OV-TM) like units decreases the coulomb repulsion energy therefore the creation of OV is likely to be optimized in such a way that most of the Mn$^{3+}$ ions will be paired as Mn$^{3+}$-OV-Mn$^{3+}$ and thus localizing the associated e$_g$-charge of 3d$_{3z2-r2}$ orbitals about the OV site. Since the site for OV will be randomly selected along X,Y,Z directions of the octahedral perovskite lattice the orientation of (Mn$^{3+}$-OV- Mn$^{3+}$) will be equally probable along these three principle symmetry and therefore the over all symmetry of the defect averaged long range phase will remain octahedral (i.e. cubic). Now since the effective concentration of 3d$_{3z2-r2}$ orbitals capable of creating orbital polarization along c-axis has decreased quite a lot (nearly vanished) there will not be any CJTD and hence there will no cubic-tetragonal phase transformation for the δ=0.14 case. Further since OV causes local symmetry lowering at some of the Mn-sites from octahedral (O$_h$) to C$_{4v}$, which is non-centrosymmetric, local spin canting of Mn moments will now be favored. These localized canted spins are expected to give extra moments above the corresponding paramagnetic background. That is what has been observed as higher moment at 350K in the 7T M-T for the off-stoichiometric sample as compared to stoichiometric one. The overall cubic symmetry may favor [1] G-type AFM phase in whose matrix the local canted spins will persist as defects. The observed broad kink like feature in the M-T corresponding to off-stoichiometric sample may be due to weak G-type AFM transition.

**Conclusion:**

The oxygen stoichiometric $La_{0.15}Sr_{0.85}MnO_3$ under goes a coupled cubic-tetragonal, semiconducting-insulating and C-type AFM phase-transition, whereas the off-stoichiometric one does not show any such transition and remain cubic with probably G-type AFM phase down to 80K. It is thus concluded that oxygen off-stoichiometry in wide band $La_{0.15}Sr_{0.85}MnO_3$ leads to complete suppression of the cooperative JT-distortion, which is necessary for coupled cubic-tetragonal and C-type AFM phase-transitions. The weak-ferromagnetism appears to be related with A-site substitutional disorder of La and Sr. Oxygen vacancy induced local spin canting in off-stoichiometric sample has made it important for the study of possible magnetoelectric-multiferroic properties. The current results indicate towards the importance and potential of oxygen-stoichiometry-control for the structural and physical property engineering of wide band manganite $La_{0.15}Sr_{0.85}MnO_3$ and possibly in general for all transition metal perovskites.

**Acknowledgment**

Authors gratefully acknowledge Dr.P.Chaddah, and Prof.A.Gupta, for their encouragement and interest in the work. Aga Shahee would like to acknowledge CSIR-India for financial support as an SRF.

**Figures and figure captions:**

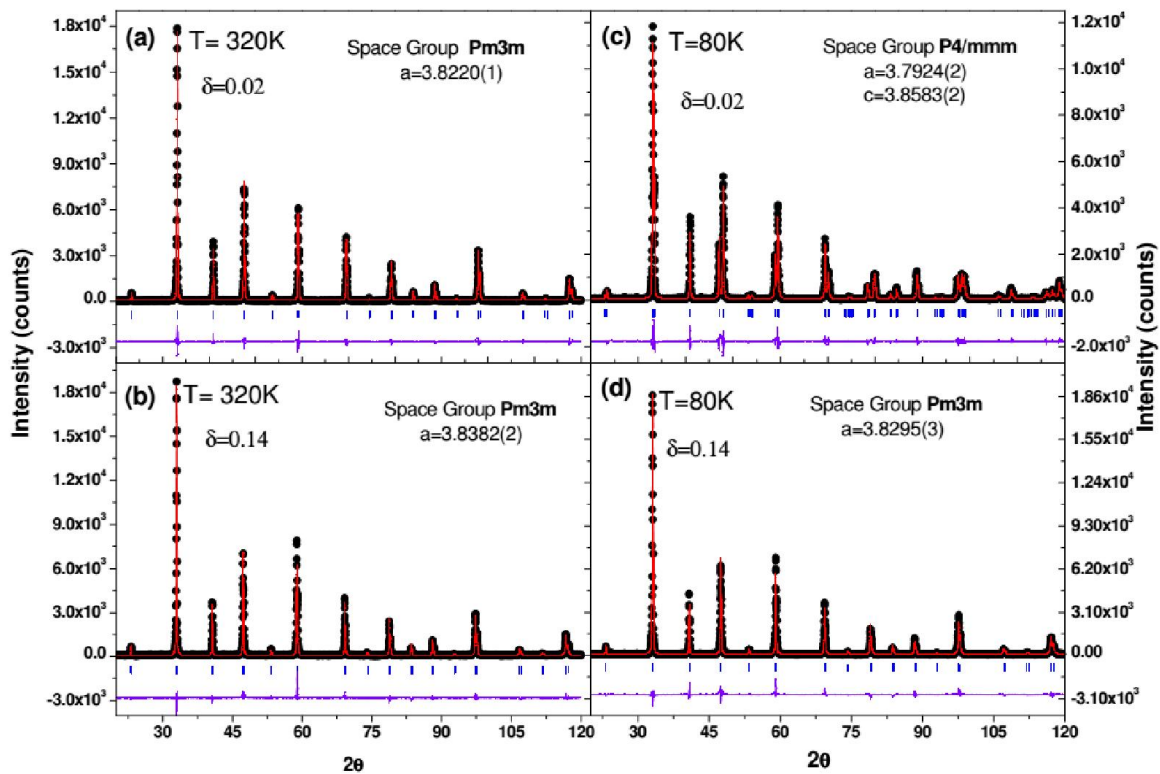

**Fig.1.** Rietveld refined XRD patterns of $La_{0.15}Sr_{0.85}MnO_{2.98}$ ($\delta=0.02$) and $La_{0.15}Sr_{0.85}MnO_{2.86}$ (i.e. $\delta=0.14$) taken at 320K and 80K. It shows that while $La_{0.15}Sr_{0.85}MnO_{2.98}$ (a,b) transforms from cubic (Pm-3m) to tetragonal (P4/mmm) (c/a=1.017) during cooling the $La_{0.15}Sr_{0.85}MnO_{2.86}$ (c,d) remains cubic down to 80K.

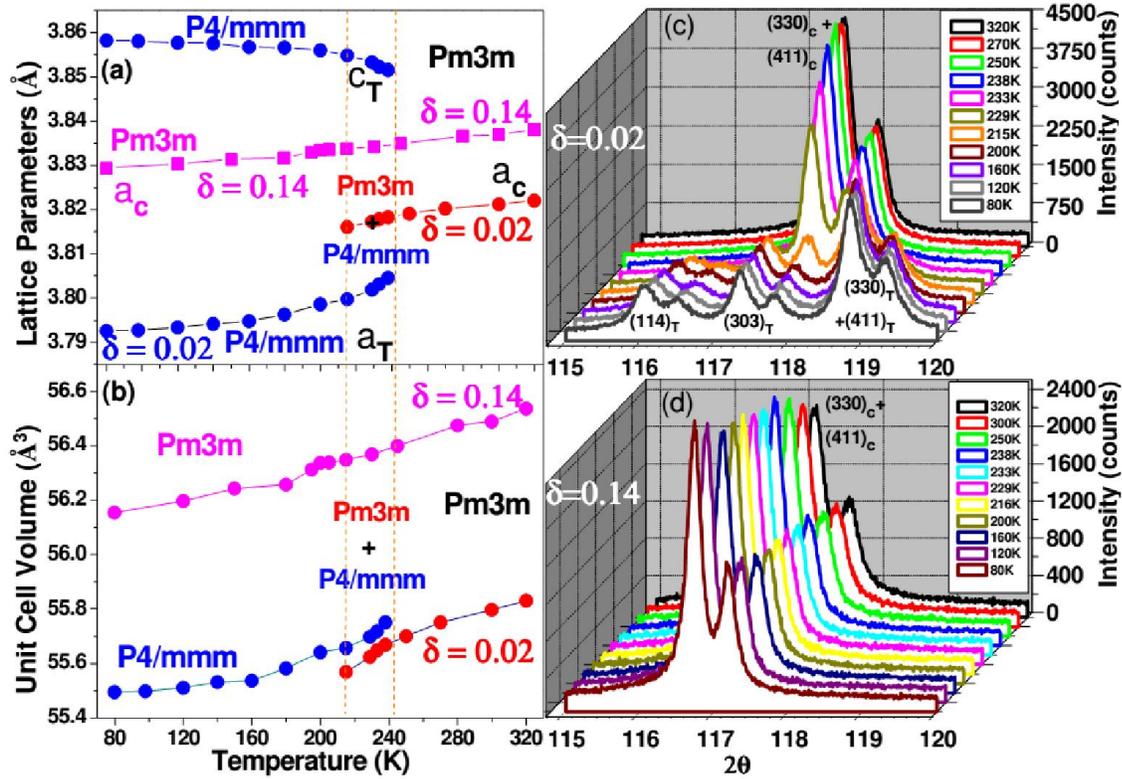

**Fig.2.** Temperature variation of (a) lattice-parameters and (b) and unit-cell volumes of $La_{0.15}Sr_{0.85}MnO_{2.98}$ and $La_{0.15}Sr_{0.85}MnO_{2.86}$ down to 80K. Splitting of degenerate reflections (330)&(411) into (114), (303), (411) and (330) due to cubic-tetragonal transformation of $La_{0.15}Sr_{0.85}MnO_{2.98}$ and its complete absence for $La_{0.15}Sr_{0.85}MnO_{2.86}$ is vividly clear from (c) and (d).

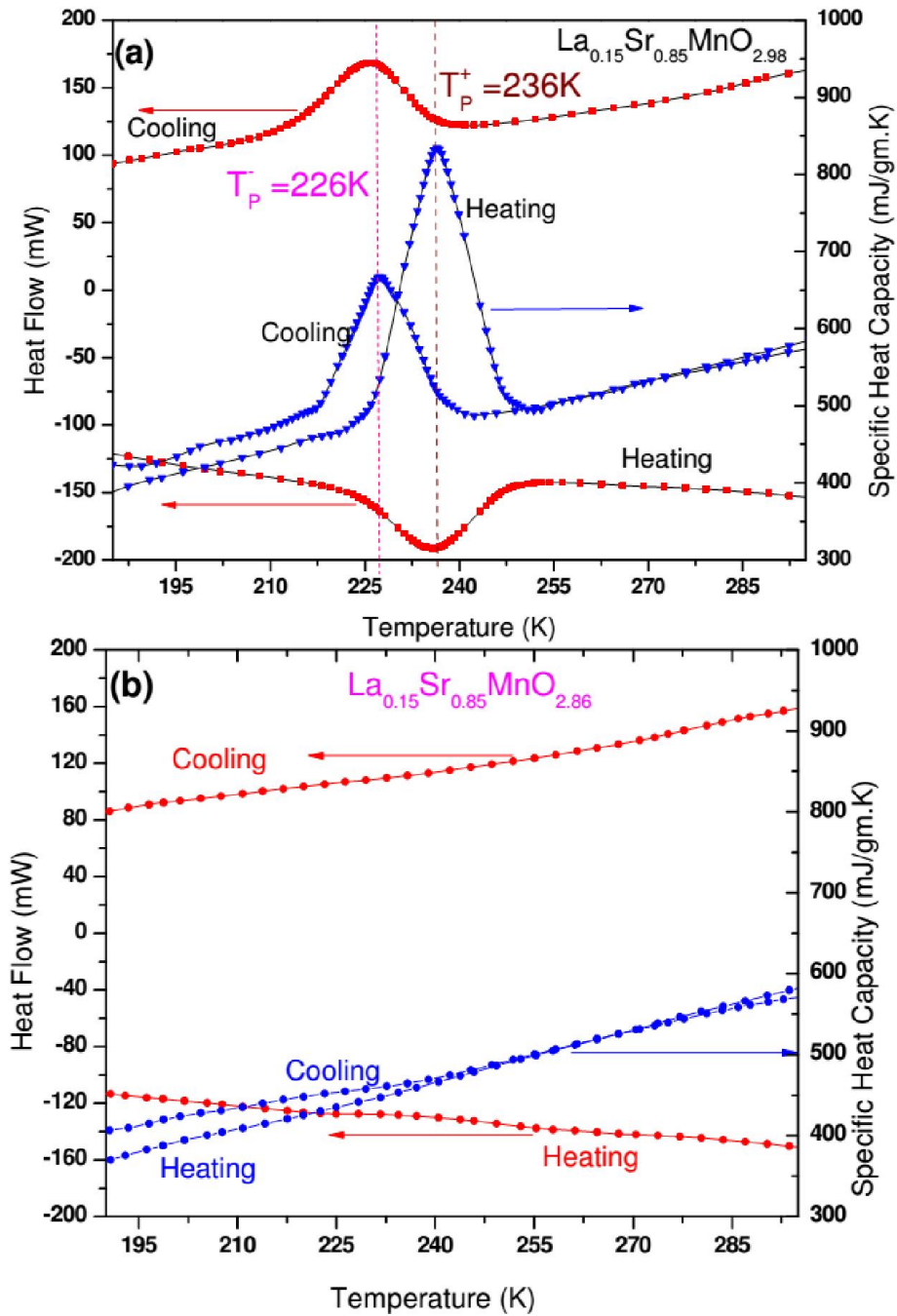

**Fig.3**. MDSC scans showing total heat-flow and $C_p$. Occurrence of peak in $C_p$-T and the corresponding exothermic/endothermic peaks in (a) shows latent-heat evolution and hence first-order nature of the cubic-tetragonal and C-type AFM transitions in $La_{0.15}Sr_{0.85}MnO_{2.98}$, while absence of these features in (b) shows thermodynamic absence of coupled structural and magnetic phase-transitions in $La_{0.15}Sr_{0.85}MnO_{2.86}$.

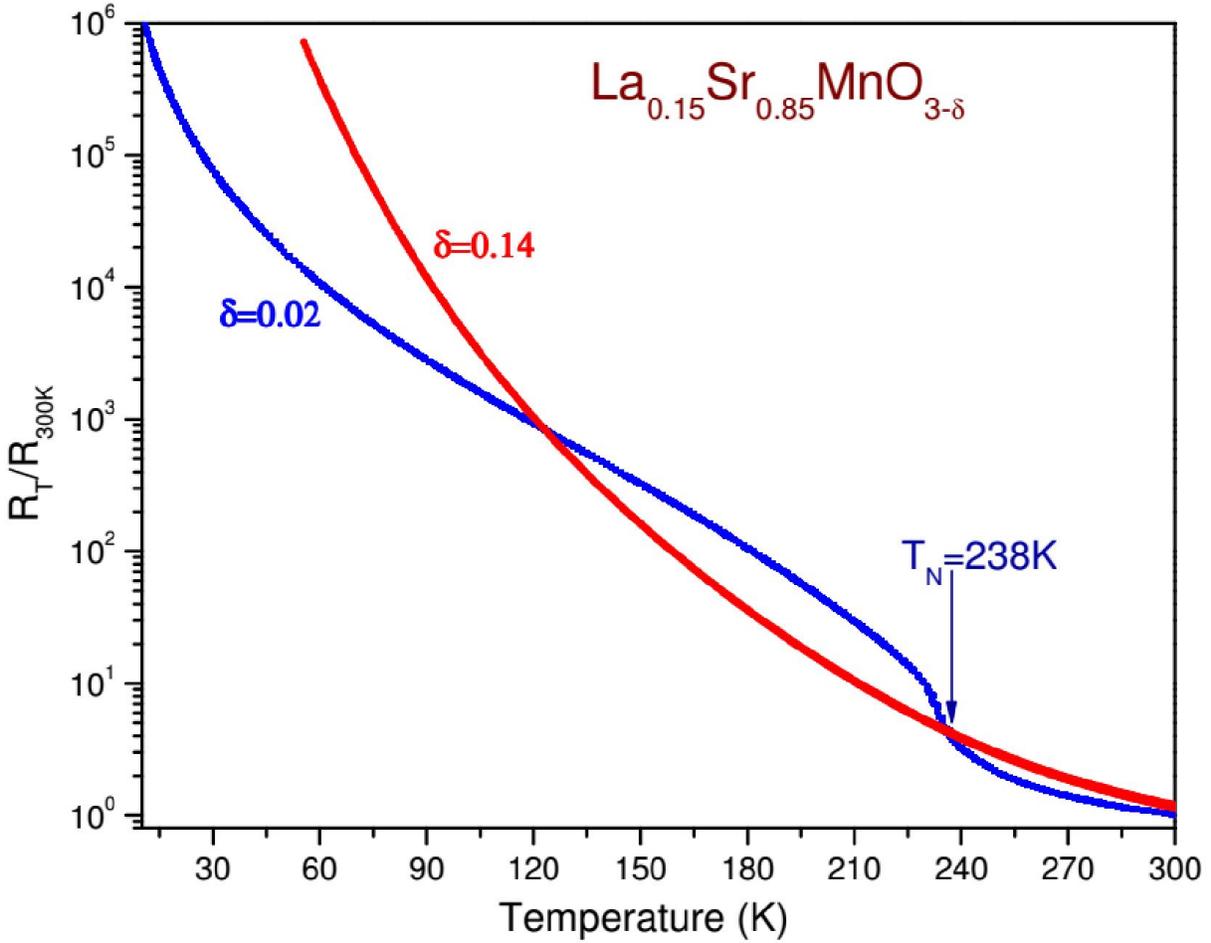

**Fig.4.** $R_T/R_{300}$ -T variation for $La_{0.15}Sr_{0.85}MnO_{2.98}$ and $La_{0.15}Sr_{0.85}MnO_{2.86}$. $La_{0.15}Sr_{0.85}MnO_{2.98}$ shows a step like feature in $R_T/R_{300}$-T, which corresponds to cubic–tetragonal and C-type AFM phase-transitions. Complete absence of this step like feature is observed for the $La_{0.15}Sr_{0.85}MnO_{2.86}$.

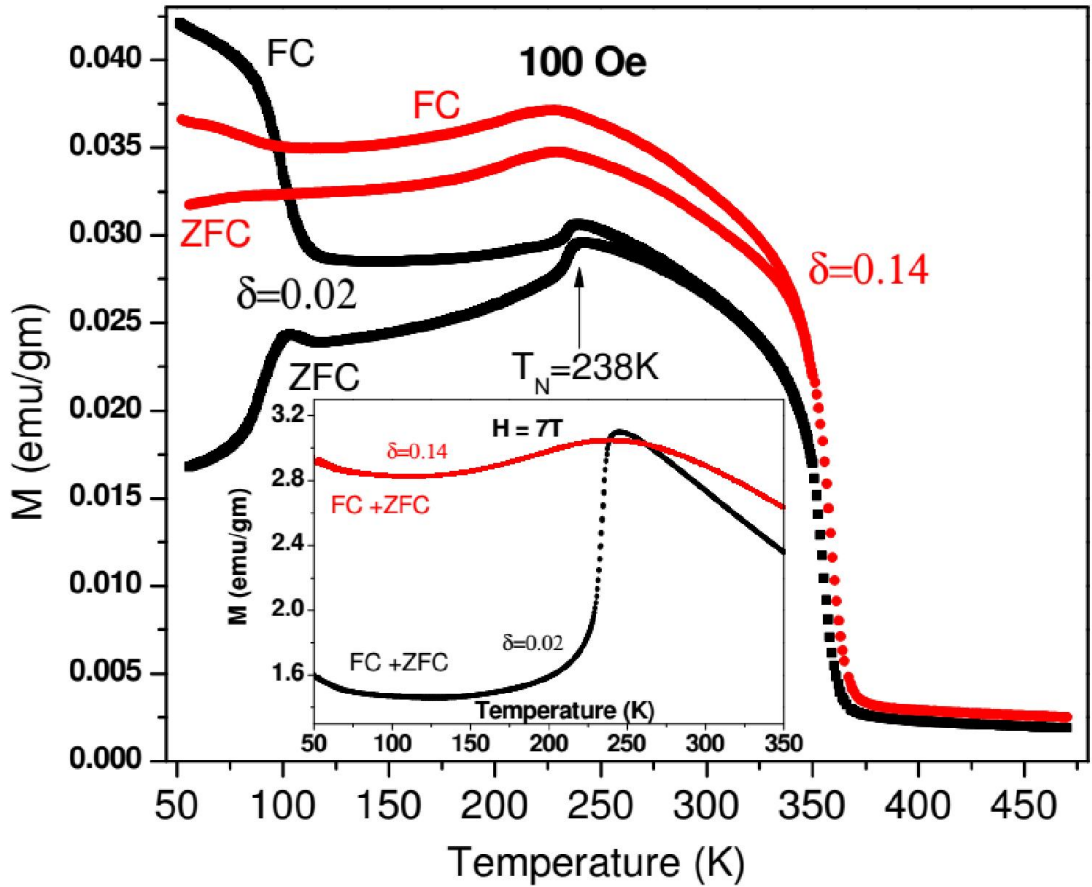

**Fig.5**. 100Oe and 7T ZFC and FC M-T data for $La_{0.15}Sr_{0.85}MnO_{2.98}$ ($\delta$=0.02) and $La_{0.15}Sr_{0.85}MnO_{2.86}$ ($\delta$=0.14). The weak-ferromagnetic like transition is observed for both the samples. A sharp-drop in M-T at ~238K corresponding to $La_{0.15}Sr_{0.85}MnO_{2.98}$ is due to C-type AFM transition, which gets suppressed for $La_{0.15}Sr_{0.85}MnO_{2.86}$.